\documentclass[cup7b]{cupbook}

\usepackage{amsmath}
\usepackage{amssymb}
\usepackage{eufrak}
\usepackage{verbatim}
\usepackage{epsfig}

\newtheorem{theorem}{Theorem}
\def\beq{\begin{equation}}
\def\eeq{\end{equation}}
\def\beqn{\begin{equation*}}
\def\eeqn{\end{equation*}}

\def\aa{\alpha}
\def\bb{\beta}

\def\dd{\delta}
\def\ee{\epsilon}

\def\oo{\omega}
\def\CC{\Gamma}

\def\zz{\zeta}

\def\be{{\bf e}}
\def\bl{{\bf l}}

\def\bu{{\bf u}}
\def\bv{{\bf v}}

\def\bK{{\bf K}}

\def\bKa{\underset{1}{\bK}}
\def\bKb{\underset{2}{\bK}}

\def\EE{\mathfrak{E}}
\def\Tb{{\bf\Theta}}

\def\Gb{{\bf\CC}}
\def\half{\tfrac{1}{2}}
\def\ra{\quad\rightarrow\quad}

\def\bk{\boldsymbol{k}}
\def\bl{\boldsymbol{l}}
\def\bm{\boldsymbol{m}}

\def\EE{\mathfrak{E}}

\def\Tb{{\bf\Theta}}
\def\Gb{{\bf\CC}}
\def\half{\tfrac{1}{2}}

\def\Ppmm<#1x>{{\hbox{${\lower3.8pt\hbox{$\bf P$}}
\atop{\smash{\raise0.2pt\hbox{$\scriptscriptstyle #1$}}}$}}}

\def\dr{\partial_r}
\def\dz{\partial_\zz}
\def\dzb{\partial_{\bar \zz}}
\def\prt<#1>{\frac{\partial}{\partial #1}}
\def\prtt<#1,#2>{\frac{\partial #1}{\partial #2}}

\def\Ppmm<#1,#2>{{\hbox{${\lower3.8pt\hbox{$\bf #2$}}
\atop{\smash{\raise0.2pt\hbox{$\scriptscriptstyle #1$}}}$}}}

\begin{document}
\author[R. P. Kerr]{\bf\Large Roy Patrick Kerr
\and {\it \normalsize ICRANet, Piazzale della Repubblica 10,
I-65122 Pescara, Italy,}\\
{\rm and} {\it \normalsize University of Canterbury, Christchurch,
New Zealand.}}

\chapter[The Kerr and Kerr--Schild metrics]{Discovering the Kerr and
Kerr--Schild metrics}

\section{Introduction}

The story of this metric begins with a paper by Alexei Zinovievich
Petrov (1954) where the simultaneous invariants and canonical forms for
the metric and conformal tensor are calculated at a general point in an
Einstein space. This paper took a
while to be appreciated in the West, probably because the Kazan
State University journal was not readily available, but Felix
Pirani (1957) used it as the foundation of an article on gravitational
radiation theory. He analyzed gravitational shock waves, calculated the
possible jumps in the Riemann tensor across the wave fronts, and
related these to the Petrov types.

I was a graduate student at Cambridge, from 1955 to 1958. In my
last year I was invited to attend the relativity seminars at
Kings College in London, including one by Felix Pirani on his 1957
paper. At the time I thought that he was stretching when he
proposed that radiation was type N, and I said so, a rather stupid
thing for a graduate student with no real supervisor to do\footnote{My
nominal supervisor was a particle physicist who had no interest in general
relativity.}. It seemed obvious that a superposition of type N solutions
would not itself be type N, and that gravitational waves near a
macroscopic body would be of general type, not Type N.

Perhaps I did Felix an injustice. His conclusions may have been
oversimplified but his paper had some very positive consequences.
Andrzej Trautman computed the asymptotic properties of the Weyl
tensor for outgoing radiation by generalizing Sommerfeld's work on
electromagnetic radiation, confirming that the far field is
Type~N. Bondi, M.G.J. van der Burg and Metzner (1962) then introduced
appropriate null coordinates to study gravitational radiation in
the far zone and related this to the results of Petrov and Pirani.

In 1958 I went to Syracuse University as a research associate of
Peter Bergman. While there I was invited to join Joshua Goldberg
at the Aeronautical Research Laboratory in Dayton
Ohio\footnote{There is a claim spread on internet that we were employed to
develop an antigravity engine to power spaceships. This is rubbish! The
main reason why the The US Air Force had created a General Relativity
section was probably to show the navy that they could also do pure
research. The only real use that the USAF
made of us was when some crackpot sent them a proposal for
antigravity or for converting rotary motion inside a spaceship to
a translational driving system. These proposals typically used
Newton's equations to prove non-conservation of momentum for some
classical system.}. There was another relativist at the lab, Dr Joseph
Schell, who had studied Einstein's unified field theory under
Vaclav Hlavaty. Josh was about to go on study leave to Europe for
a few months, and did not want to leave Joe by himself.

Before he left, Josh and I became interested in the new methods
that were entering general relativity from differential geometry
at that time. We did not have a copy of Petrov's paper so our
first project was to re-derive his classification using projective
geometry, something which was being done by many other people
throughout the world at that time. In each empty Einstein space,
$\EE$, the conformal tensor determines four null ``eigenvectors'' at
each point. The metric is called algebraically special
(AS)\footnote{The term algebraically degenerate is sometimes used
instead.} if two of these eigenvectors coincide. This vector is
then called a principal null vector (PNV) and the field of these is called
a principal null ``congruence''.

After this, we used a tetrad formulation to study vacuum Einstein
spaces with degenerate holonomy groups (Goldberg and Kerr (1961),
Kerr and Goldberg (1961)). The tetrad used consisted of two null
vectors and two {\it real} orthogonal space-like vectors,
\beqn
ds^2 = (\oo^1)^2 + (\oo^2)^2 + 2\oo^3\oo^4.
\eeqn
We proved that the holonomy group must be an even
dimensional subgroup of the Lorentz group at each point, and
that if its dimension is less than six then coordinates can be
chosen so that the metric has the following form:
\beqn
ds^2 = dx^2 + dy^2 + 2du(dv + \rho dx + \tfrac{1}{2}(\omega
-\rho_{,x}v)du),
\eeqn
where both $\rho$ and $\omega$ are independent of $v$\footnote{The simple
way that the coordinate v appears was to prove typical of these
algebraically special metrics.}, and
\begin{gather*}
\rho_{,xx} + \rho_{,yy} = 0\\
\omega_{,xx} + \omega_{,yy} = 2\rho_{,ux} - 2 \rho\rho_{,xx} -
(\rho_{,x})^2 + (\rho_{,y})^2
\end{gather*}
This coordinate system was not quite uniquely defined. If $\rho$ is
bilinear in $x$ and $y$ then it can be transformed to zero, giving
the well-known plane-fronted wave solutions. These are type N, and
have a two-dimensional holonomy groups. The more general metrics are
type III with four-dimensional holonomy groups.

In September 1961 Joshua joined Hermann Bondi, Andrzej Trautman,
Ray Sachs and others at King's College in London. By this time it was
well known that all such AS spaces possess a null congruence whose
vectors are both geodesic and shearfree. These are the degenerate
``eigenvectors'' of the conformal tensor at each point. Andrzej
suggested to Josh and Ray how they might prove the converse. This
led to the celebrated Goldberg-Sachs theorem (see Goldberg and
Sachs (1962)):

\begin{theorem}A vacuum metric is algebraically special if and
only if it contains a geodesic and shearfree null congruence.
\end{theorem}

Either properties of the congruence, being geodesic and shear-free,
or a property of the conformal tensor, algebraically
degeneracy, could be considered fundamental with the others
following from the Goldberg-Sachs theorem. It is
likely that most thought that the algebra was fundamental,
but I believe that Ivor Robinson and Andrzej Trautman (1962)
were correct when they emphasized the properties of the congruence
instead. They showed that for any Einstein space with a shear-free
null congruence which is also hypersurface orthogonal there are
coordinates for which
\beqn ds^2= 2r^2P^{-2} d\zz d \bar \zz - 2dudr - (\Delta \ln P
-2r(\ln P)_{,u} -2m(u)/r)du^2,
\eeqn
where $\zz$ is a complex coordinate, $ = (x+iy)/\sqrt{2}$, say, so that
\beqn
2d\zz d \bar \zz = dx^2+dy^2.
\eeqn
The one remaining field equation is,
\beqn
\Delta\Delta(\ln P) + 12m(\ln P)_{,u} - 4m_{,u} = 0,\quad
\Delta = 2P^2 \dz \dzb.
\eeqn

The PNV\footnote{The letter $\bk$ will be used throughout this
article to denote the PNV.} is $\bk = k^\mu\partial_\mu = \dr$,
where $r$ is an affine parameter along the rays. The corresponding
differential form is $k = k_\mu dx^\mu = du$, so that $\bk$ is the
normal to the surfaces of constant $u$. The coordinate $u$ is a
retarded time, the surfaces of constant $r,u$ are distorted
spheres with metric $ds^2= 2r^2P^{-2} d\zz d \bar \zz$ and the
parameter $m(u)$ is loosely connected with the
system's mass. This gives the complete solution to the
Robinson--Trautman problem\footnote{In the study of exact
solutions, ``solving'' a problem usually means introducing a useful
coordinate system, solving the easier Einstein equations and
replacing the ten components of the metric tensor
with a smaller number of functions, preferably of less than four
variables. These will then have to satisfy a residual
set of differential equations, the harder ones, which usually have no
known complete solution. For example, the remaining field equation
for the Robinson--Trautman metrics is
highly nonlinear and has no general solution.}.
%\begin{comment}
\begin{figure}
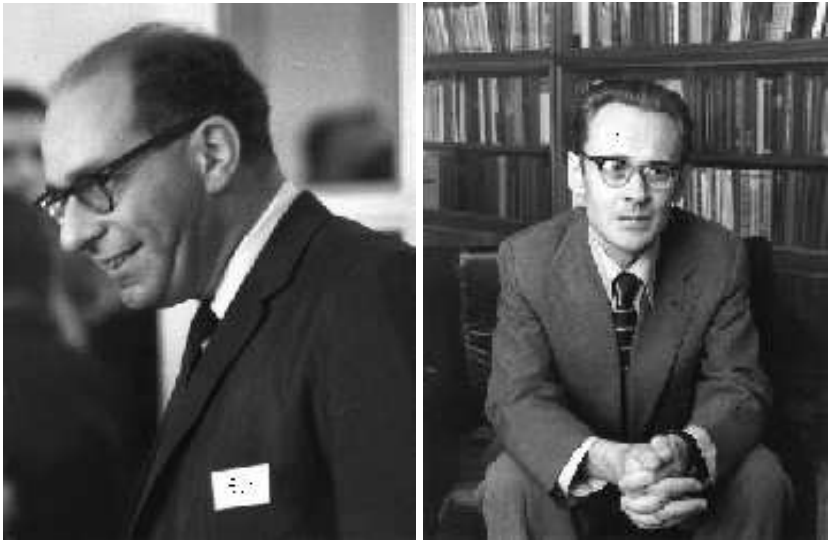

\begin{center}
\includegraphics[width=.5\hsize,clip]{RobinsonPic.eps}
\includegraphics[width=.49\hsize,clip]{TrautmanPic.eps}
\caption{ Ivor Robinson and Andrzej Trautman constructed all Einstein spaces possessing a hypersurface orthogonal shearfree congruence. Whereas Bondi and his colleagues were looking at spaces with these properties asymptotically, far from any sources, Robinson and Trautman went a step further, constructing exact solutions. (Images courtesy of Andrzej Trautman and the photographer, Marek Holzman)}\label{Rob}
\end{center}
\end{figure}

%\end{comment}

In 1962 Goldberg and myself attended a month-long meeting in Santa
Barbara. It was designed to get mathematicians and relativists
talking to each other. Perhaps the physicists learned a lot about
more modern mathematical techniques, but I doubt that the
geometers learned much from the relativists. All that aside, I met
Alfred Schild at this conference. He had just persuaded the Texas state
legislators to finance a Center for Relativity at the University
of Texas, and had arranged for an outstanding group of relativists
to join. These included Roger Penrose and Ray Sachs, but neither
could come immediately and so I was invited to visit for the
62-63 academic year.

After Santa Barbara, Goldberg and myself flew to a conference held
at Jablonna near Warsaw. This was the third precursor to the
triennial meetings of the GRG society and therefore it could be
called GR3. Robinson and Trautman (1964) presented a
paper on {\it ``Exact Degenerate Solutions''} at this conference.
They spoke about their well-known solution and also showed that when
the rays are not hypersurface orthogonal coordinates can be chosen so that
\beqn
ds^2 = -P^2[(d\xi - a k)^2 + (d\eta - b k)^2]
+2 d\rho k + c k^2,
\eeqn
where, as usual, $k$ is the PNV. Its components, $k_\aa$, are
independent of $\rho$, but $a, b, c$ and $P$ may be functions of all
four coordinates.

\begin{figure}
\begin{center}
\includegraphics[scale= .6]{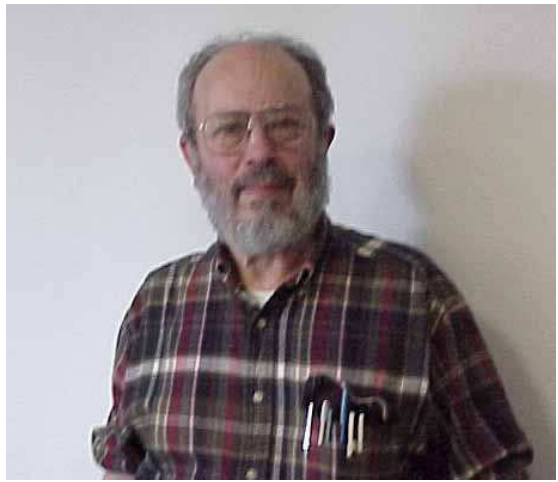}
\caption{Ezra T Newman, with T. Unti and L.A. Tambourino, studied the field equations for diverging and rotating algebraically special Einstein spaces.}\label{New}
\end{center}
\end{figure}

I was playing around with the structure of the Einstein equations
during 1962, using the new (to physicists) methods of tetrads and
differential forms. I had written out the equations for the
curvature using a complex null tetrad and self-dual bivectors, and
then studied their integrability conditions. In particular, I
was interested in the same problem that Robinson and Trautman were
investigating where $k$ was not a gradient, i.e twisting, but there
was a major road block in my way. Alan
Thompson had also come to Austin that year and was also interested
in these methods. Although there seemed to be no reason why there
should not be many algebraically special spaces, Alan kept quoting
a result from a preprint of a paper by Newman, Tambourino and Unti (1963) in
which they had ``proved'' that the only possible space with a
diverging and rotating PNV is NUT space, a one parameter
generalization of the Schwarzschild metric. They derived this
result using the new Newman--Penrose spinor formalism (N--P). Their
equations were essentially the same as those
obtained by people such as myself using self-dual bivectors: only
the names are different. I could not understand how the equations
that I was studying could possibly lead to their claimed result,
but could only presume it must be so since I did not have a copy
of their paper.

In the spring of 1963 Alan obtained a preprint of this paper and loaned it
to me. I thumbed through it quickly, trying to see where their hunt for
solutions had died. The N--P formalism assigns a different Greek
letter to each component of the connection, so I did not try to
read it carefully, just rushed ahead until I found what appeared
to be the key equation,
\beq\label{nsum}
\tfrac{1}{3}(n_1 + n_2 + n_3)a^2 = 0,
\eeq
where the $n_i$ were all small integers, between -4
and +4. Their sum was not zero so this gave $a = 0$. I had no
idea what $a$ represented, but its vanishing seemed to be
disastrous and so I looked more carefully to see where this
equation was coming from. Three of the previous equations, each
involving first derivatives of some of the connection components, had
been differentiated and then added together. All the second derivatives
cancelled identically and most of the other terms were eliminated using
other N--P equations, leaving (\ref{nsum}).

The mistake that Newman~{et al.} made was that they did not notice that
they were simply recalculating
one component of the Bianchi identities by adding together the
appropriate derivatives of three of their curvature equations, and
then simplifying the result by using some of their other equations,
undifferentiated. The result should have agreed with one of their derived
Bianchi identities involving derivatives of the components of the conformal
tensor, the $\Psi_i$ functions, giving
\beq\label{nsum:1}
n_1 + n_2 + n_3 \equiv 0.
\eeq
In effect, they rediscovered one component of the identities, but with
numerical errors. The real fault was the way the N--P
formalism confuses the Bianchi identities with the derived
equations involving derivatives of the $\Psi_i$ variables.

Alan Thompson and myself were living in adjoining apartments, so I
dashed next door and told him that their result was incorrect. Although it was
unnecessary, we recalculated the first of the three terms, $n_1$, obtained
a different result to the one in the preprint, and verified that
(\ref{nsum:1}) was now satisfied. Once this blockage was out of the way,
I was then able to continue with what I had been doing and derive the
metric and field equations for twisting algebraically special spaces. The
coordinates I constructed turned out to be essentially the same as
the ones given in Robinson and Trautman (1962). This shows that
they are the ``natural'' coordinates for this problem since the
methods used by them were very different to those used by me. Ivor
loathed the use of such things as N--P or rotation coefficients,
and Andrzej and he had a nice way of proving the existence of their
canonical complex coordinates $\zz$ and $\bar\zz$. I found this
same result from one of the Cartan equations, as will be shown in
the next section, but I have no doubt that their method is more
elegant. Ivor explained it to me on more than one occasion, but
unfortunately I never understood what he was saying\footnote{While
writing this article I read their 1962 paper and finally
understood how they derived their coordinates. It only took me 45
years.}!

At this point I presented the results at a monthly Relativity
conference held at the Steven's Institute in Hoboken, N.J. When I
told Ted Newman that (\ref{nsum}) should have been identically
zero, he said that they knew that $n_1$ was incorrect, but that
the value for $n_2$ given in the preprint was a misprint and so
(\ref{nsum:1}) was still not satisfied. I replied that since the
sum had to be zero the final term, $n_3$ must also be incorrect.
Alan and I recalculated it that evening, confirming that
(\ref{nsum:1}) was satisfied\footnote{Robinson and Trautman (1962)
also doubted the original claim by Newman~et~al.\ since they knew
that the linearized equations had many solutions.}.

\section{Discovery of the Kerr metric}

When I realized that the attempt by Newman et al.\ to find all
rotating AS spaces had foundered and that Robinson and Trautman
appeared to have stopped with the static ones, I rushed headlong
into the search for these metrics.

Why was the problem so interesting to me?
Schwarzschild, by far the most significant physical solution known at
that time, has an event horizon. A spherically symmetric star that
collapses inside this is forever lost to us, but it was not known
whether angular momentum could stop this collapse to a black hole.
Unfortunately, there was no known metric for a rotating star. Schwarzschild
was an example of the Robinson--Trautman metrics, none of which
could contain a rotating source as they were all hypersurface orthogonal.
Many had tried to solve the Einstein equations assuming a stationary
and axially symmetric metric, but none had
succeeded in finding any physically significant rotating solutions. The
equations for such metrics are complicated nonlinear PDEs in two variables.
What was needed was some extra condition that would reduce these to ODEs,
and this might be the assumption that the metric is AS.

The notation used in the rest of this paper is fairly standard. There were
two competing formalisms being used around 1960, complex tetrads and spinors.
I used the former\footnote{Robinson and Trautman also had a fairly
natural complex tetrad approach.}, Newman et al.\ the latter. The
derived equations are essentially identical, but each approach has
some advantages. Spinors make the the Petrov classification
trivial once it has been shown that a tensor with the symmetries
of the conformal tensor corresponds to a completely symmetric
spinor, $\Psi_{ABCD}$. The standard notation for the components of
this tensor is
\beqn \Psi_0 = \Psi_{0000}, \quad \Psi_1 = \Psi_{0001},
\;\dots\; \Psi_4 = \Psi_{1111}. \eeqn
If $\zz^A$ is an arbitrary spinor then the equation
\beqn
\Psi_{ABCD}\zz^A\zz^B\zz^C\zz^D = 0
\eeqn
is a homogeneous quartic equation with four complex roots,
$\zz^A_i$. The related real null vectors, $Z^{\aa\dot \aa}_i =
\zz^\aa_i \zz^{\dot\aa}_i$, are the four PNVs of Petrov. The
spinor $\zz^\aa = \dd^\aa _0$ gives a PNV if $\Psi_0 = 0$. It is a
repeated root and therefore it is the principal null vector of an
AS spacetime if $\Psi_1 = 0$ as well.

The Kerr (1963) letter presented the main results of my
calculations but gave few details\footnote{I spent many years
trying to write up this research but, unfortunately, I could never
decide whether to use spinors or a complex tetrad, and thus it did
not get written up until Kerr and Debney
(1969). George Debney also collaborated with Alfred Schild and myself
on the Kerr--Schild metrics in Debney et al.\ (1970).}. The methods that
I used to solve the equations for AS spaces are essentially those used
in ``Exact Solutions'' by Stephani et al.\ (2003), culminating in their
equation (27.27). I will try to use the same notation as in that
book since it is almost identical to the one I used in 1963, but I
may get some of the signs wrong. Beware!

Suppose that $(\be_a) = (\bm, \bar{\bm}, \bl,\bk)$ is a null
tetrad, i.e.,a set of four null vectors where the last two are
real and the first two are complex conjugates. The corresponding
dual forms are $(\oo^a) = (\bar m , m, -k, -l)$ and the metric is
\beq\label{E:ds2}
ds^2 = 2(m \bar m - k l) = 2(\oo^1 \oo^2 - \oo^3 \oo^4).\footnote{I
personally hate the minus sign in this expression and did not use it in
1963, but it seems to have become standard. By the time I finish this
article I am sure that I will wish I had stuck with all positive signs!}
\eeq
The vector $\bk$ is a PNV and so its direction is uniquely defined,
but the other directions are not. The form of the metric tensor in
\eqref{E:ds2} is invariant under a combination of a null rotation
($B$) about $\bk$, a rotation ($C$) in the $\bm \wedge \bar \bm$
plane and a Lorentz transformation ($A$) in the $\bl \wedge \bk$
plane,
\begin{subequations}
\begin{align}
\bk^\prime &= \bk, \quad &\bm^\prime &= \bm + B \bk,
& \bl^\prime &= \bl + B\bar\bm + \bar B \bm + B \bar B
\bk,\\
\bk^\prime &= \bk, &\bm^\prime &= e^{iC}\bm, &\bl^\prime
&=\bl, \label{E:Ctrans}\\
\bk^\prime &= A\bk, &\bm^\prime &=\bm, & \bl^\prime &=
A^{-1}\bl. \label{E:Atrans}
\end{align}
\end{subequations}
The most important connection form is
\beqn
\Gb_{41} = \Gamma_{41a}\oo^a = m^\aa k_{\aa;\bb}dx^\bb
\eeqn
The optical scalars of Ray Sachs for $\bk$ are just the components
of this form with respect to the $\oo^a$.
\begin{align*}
\sigma &= \CC_{411} = {\rm shear.}\\
\rho &= \CC_{412} = {\rm complex\ divergence,}\\
\kappa &= \CC_{414} = {\rm geodesy,}
\end{align*}

The fourth component, $\CC_{413}$, is not invariant under
a null rotation about $ \bk$,
\beqn
\CC_{413}^\prime = \CC_{413} + B\rho,
\eeqn
and has no real geometric significance. It can be set to zero
using an appropriate null rotation. Also, since $\bk$ is geodesic and
shearfree for AS metrics, both $\kappa$ and
$\sigma$ are zero and therefore
\beq\label{E:G41}
\Gb_{41} = \rho \oo^2.
\eeq
If we use the simplest field equations,
\beqn
R_{44} = 2R_{4142} = 0, \quad R_{41} = R_{4112} - R_{4134} =
0, \quad R_{11} = 2R_{4113} = 0,
\eeqn
a total of 5 real equations, and the fact that the metric is AS,
\beqn
\Psi_0 = - 2R_{4141} = 0, \quad 2\Psi_1 = - R_{4112} -
R_{4134}= 0,
\eeqn
then the most important of the second Cartan equations
simplifies to
\beq\label{E:dG41}
d\Gb_{41} - \Gb_{41} \wedge(\Gb_{12} +\Gb_{34} ) =
R_{41ab}\oo^a\wedge \oo^b = R_{4123}\oo^2\wedge \oo^3.
\eeq

Taking the wedge product of (\ref{E:dG41}) with $ \Gb_{41}$ and
using (\ref{E:G41}),
\beq\label{E:GdG}
\Gb_{41} \wedge d\Gb_{41} = 0.
\eeq
This was the key step in my study of these metrics but this result
was not found in quite such a simple way. At first, I stumbled around using
individual component equations rather than differential forms to
look for a useful coordinate system. It was only after I
had found this that I realized that using differential forms
from the start would have short-circuited several days analysis.

Equation (\ref{E:GdG}) is the integrability condition for the
existence of complex functions, $\zz$ and $\Pi$, such that
\beqn
\Gb_{41} = d \bar\zz / \Pi, \quad \Gb_{42} = d \zz /
\bar\Pi.
\eeqn
The two functions $\zz$ and its complex conjugate, $\bar\zz$,
will be used as (complex) coordinates. They are not quite unique
since $\zz$ can always be replaced by an arbitrary analytic
function $\Phi (\zz)$.

Using the transformations in (\ref{E:Ctrans}) and (\ref{E:Atrans}),
$$\Gb_{4^\prime1^\prime} = Ae^{iC}\Gb_{41} = Ae^{iC}d
\bar\zz / \Pi,\quad\Rightarrow\quad \Pi^\prime = A^{-1}e^{-iC}\Pi.
$$
$\Pi^\prime$ can be set to $1$ by choosing $Ae^{iC} = \Pi$, and
that is what I did in 1963, but it is also common to just use the
$C$-transformation to convert it to a real function $P$,
\beq
\Gb_{41} = \rho \oo^2 = d\bar\zz / P \quad.
\eeq
This is the derivation for two of the coordinates used in 1963.
Since $\oo^1_\aa k^\aa = 0 \rightarrow \bk( \zz)=0$, these
functions, $\zz, \bar\zz$, are constant along the PNV.

The other two coordinates were very standard and were used by most
people considering similar problems at that time. The simplest
field equation is
\beqn
R_{44} = 0 \quad \Rightarrow \quad \bk \rho = \rho_{|4} = \rho^2,
\eeqn
so that the real part of $-\rho^{-1}$ is an affine parameter along
the rays. This was the obvious choice for the third coordinate, r,
$$ \rho ^{-1} = - (r+ i \Sigma).$$
There was no clear choice for the fourth coordinate, so $u$ was chosen
so that $l^\aa u_{,\aa} = 1,\;k^\aa u_{,\aa} = 0,$ a pair of
consistent equations.

Given these four coordinates, the basis forms are
\begin{align*}
\oo^1 &= m_\aa dx^\aa = - d\zz /P \bar\rho = (r- i \Sigma)d\zz /P ,\\
\oo^2 &= \bar m_\aa dx^\aa = - d\bar\zz / P\rho = (r+i\Sigma) d\bar\zz /P,\\
\oo^3 & = k_\aa dx^\aa\;= \;\; du + Ld\zz + \bar L d\bar\zz,\\
\oo^4 & = l_\aa dx^\aa \;\;= \;\;dr + Wd\zz + \bar W d \bar \zz + H\oo^3.
\end{align*}
where $L$ is independent of $R$, and the coefficients $\Sigma$,
$W$ and $H$ have still to be determined.

When all this was substituted into the first Cartan equation,
(\ref{E:cart:a}), and (\ref{E:dG41}), the simplest component of the second
Cartan equation, (\ref{E:cart:b}), $\Sigma$ and $W$ were calculated as
functions of $L$ and its derivatives\footnote{In Kerr (1963)
$\Omega$, $D$ and $\Delta$ were used instead of $L$, $\partial$ and $\Sigma$,
but the results were the same, mutatis mutandis.},
\begin{align*}
2i\Sigma &= P^2(\bar \partial L - \partial \bar L),\quad \partial =
\partial _\zz - L \partial_u,\\
W &= - (r+ i \Sigma) L_{,u} + i \partial \Sigma.
\end{align*}
The remaining field equations, the ``hard'' ones, were more
complicated, but still fairly straightforward to calculate. Two
gave $H$ as a function of a real ``mass'' function $m(u,\zz,
\bar\zz)$ and certain functions of the higher derivatives of $P$ and
$L$\footnote{This expression for $M$ was first published by
Robinson et al.\ (1969). The corresponding expression in Kerr
(1963) is for the gauge when $P=1$. The same is true for equation
(\ref{E:kfe}c).},
\begin{align*}
H &= \half K
- r (\ln P)_{,u} - \frac{mr + M \Sigma}{r^2 +
\Sigma^2}.\\
M &= \Sigma K + P^2Re[\partial\bar\partial \Sigma -
2 \bar L_{,u} \partial \Sigma - \Sigma \partial_u
\partial\bar L] ,\\
K &= 2P^{-2} Re[\partial(\bar\partial \ln P -\bar L_{,u})],
\end{align*}
Finally, the first
derivatives of the mass function, $m$, are given by the rest of the
field equations, $R_{31}=0$ and $R_{33}=0$,
\begin{subequations}\label{E:kfe}
\begin{align}
\partial(m+iM) &= 3(m+iM)L_{,u},\\
\bar\partial(m-iM) &= 3(m-iM)\bar L_{,u} ,\\
[P^{-3} (m + iM)]_{,u} &= P[\partial + 2(\partial \ln P -
L_{,u}]\partial I,
\end{align}
\end{subequations}
where
\beq
I = \bar\partial(\bar \partial \ln P - \bar L_{,u})
+ (\bar \partial \ln P - \bar L_{,u})^2.
\eeq

As was said in Kerr (1963), there are two natural choices that can
be made to restrict the coordinates and simplify the final
results. One is to rescale $r$ so that $P=1$ and $L$ is complex,
the other is to take $L$ to be pure imaginary with $P \neq 1$. I
chose to do the first since this gives the most concise form for
$M$ and the remaining field equations. It also gives the smallest
group of permissible coordinate transformations, simplifying the
task of finding all possible Killing vectors. The results for this
gauge are
\begin{subequations}\label{E:Keq}
\begin{gather}
M = {\rm Im}(\bar\partial\bar\partial\partial L),\\
\partial(m + iM) = 3(m+iM)L_{,u},\\
\bar\partial(m-iM) = 3(m-iM)\bar L_{,u} ,\\
\partial_{u}[m - {\rm Re}(\bar\partial\bar\partial\partial L)]= |\partial_u
\partial L|^2.
\end{gather}
\end{subequations}

Since all derivatives of $m$ are given, the commutators were calculated
to see whether the field equations were completely integrable.
This gives $m$ as a function of the higher derivatives of $L$ unless
both $\Sigma_{,u}$ and $L_{,uu}$ are zero. As stated in Kerr
(1963), if these are both zero then there is a coordinate system in
which $P $ and $L$ are independent of $u$ and $m = cu+A(\zz,
\bar\zz)$, where c is a real constant. If it is zero then
the metric is independent of u and therefore stationary. The field
equations are
\begin{subequations}\label{E:withC}
\begin{align}
\nabla[&\nabla(\ln P)] = c,&\nabla =
P^2\partial^2/\partial\zz\partial\bar\zz,\\
M &= 2\Sigma \nabla(\ln P) + \nabla \Sigma, &m = cu + A(\zz, \bar\zz),\\
cL &= ( A + i M)_\zz, \qquad\Rightarrow &\nabla M = c\Sigma.\qquad\quad
\end{align}
\end{subequations}
We shall call these metrics quasi-stationary.

In Kerr (1963) I stated that the solutions of these equations include the Kerr
metric (for which $c=0$). This is true but it is not how this solution was
found. Furthermore, in spite of what many believe, its construction
did not use the Kerr--Schild ansatz.

\section{Symmetries in Algebraically Special Spaces}

The field equations, (\ref{E:kfe}) or (\ref{E:Keq}), are so
complicated that some extra assumptions were needed to reduce them
to a more manageable form. My next step in the hunt for physically
interesting solutions was to assume that the metric is stationary.
Fortunately, I had some tricks that allowed me to find all
possible Killing vectors without actually solving Killing's
equation.

The key observation is that any Killing vector generates a
1-parameter group which must be a subgroup of the group
$\mathcal{C}$ of coordinate transformations that preserve all
imposed coordinate conditions.

Suppose that $\{x^{\star a}, \oo_a^\star\}$ is another set of
coordinates and tetrad vectors that satisfy the conditions imposed
in the previous section. If we restrict our coordinates to those that
satisfy $P=1$ then $\mathcal{C}$ is the group of transformations
$x\rightarrow x^\star$ for which
\begin{align*}\label{}
\zz^\star &= \Phi(\zz),&&\oo^{1\star} = (|\Phi_\zz|/\Phi_\zz)\oo^1,\\
u^\star &= |\Phi_\zz|(u + S(\zz,\bar\zz),
&&\oo^{3\star}=|\Phi_\zz|^{-1}\oo^3,\\
r^\star &= |\Phi_\zz|^{-1} r,&&\oo^{4\star} =|\Phi_\zz|\oo^4,
\end{align*}
and the basic metric functions, $L^\star$ and $m^\star$, are given by
\begin{subequations}\label{E:mL}
\begin{gather}
L^\star = (|\Phi_\zz|/\Phi_\zz)[L - S_\zz -
\half(\Phi_{\zz\zz}/\Phi_\zz)(u + S(\zz,\bar\zz)],\\
m^\star = |\Phi_\zz|^{-3}m.\quad\quad
\end{gather}
\end{subequations}

Let $\mathcal{S}$ be the identity component of the group of
symmetries of a space. If we interpret these as coordinate
transformations, rather than point transformations, then it is the
set of transformations $x\rightarrow x^\star$ for which
\beqn
g^\star_{\aa\bb}(x^\star) = g_{\aa\bb}(x^\star).
\eeqn

It can be shown that $\mathcal{S}$ is precisely the subgroup of
$\mathcal{C}$ for which
\beqn
m^\star(x^\star) = m(x^\star), \quad L^\star(x^\star) =
L(x^\star).\footnote{Note that this implies that all derivatives of
these functions are also invariant, and so $g_{\aa\bb}$ itself is invariant.}
\eeqn
Suppose that $x\rightarrow x^\star(x,t)$ is a 1-parameter group of
motions,
\begin{align*}\label{}
\zz^\star &= \psi(\zz;t),\\
u^\star &= |\psi_\zz|(u + T(\zz,\bar\zz;t),\\
r^\star &= |\psi_\zz|^{-1} r.
\end{align*}
Since $x^\star(x;0) = x$, the initial values of $\psi$ and $T$ are
\beqn
\psi(\zz;0) = \zz, \quad T(\zz,\bar\zz;0) = 0.
\eeqn
The corresponding infinitesimal transformation, $\bK =
K^\mu\partial/\partial x^\mu$ is
\beqn
K^\mu = \left[\frac{\partial x^{\star\mu}}{\partial t}\right]_{t=0}.
\eeqn
If we define
\beqn
\aa(\zz) = \left[\frac{\partial\psi}{\partial t}\right]_{t=0}, \quad
V(\zz,\bar\zz) = \left[\frac{\partial T}{\partial t}\right]_{t=0},
\eeqn
then the infinitesimal transformation is
\beq\label{}
\bK = \aa\partial_\zz + \bar\aa \partial_{\bar\zz} +
Re(\aa_\zz)(u\partial_u - r\partial_r) + V\partial_u.
\eeq
Replacing $\Phi(\zz)$ with $\psi(\zz; t)$ in (\ref{E:mL}),
differentiating this w.r.t. t, and using the initial values for
$\psi$ and $T$, $\bK$ is a Killing vector if and only if
\begin{gather*}
V_\zz + \half\aa_{\zz\zz}r + \bK L + \half(\aa_\zz -
\bar\aa_{\bar\zz})L = 0,\\
\bK m + 3 Re(\aa_\zz) m = 0.
\end{gather*}
The transformation rules for $\bK$ under an element $(\Phi, S)$ of
$\mathcal{C}$ are
\beqn
\aa^\star = \Phi_\zz \aa, \quad
V^\star = |\Phi_\zz|[V - Re(\aa_\zz)S + \bK S].
\eeqn

Since $\aa$ is itself analytic, if $\aa \neq 0$ for a particular
Killing vector then, $\Phi$
can be chosen so that $\aa^\star = 1$\footnote{Or any other analytic
function of $\zz$ that one chooses}, and then $S$ so that $V^\star =
0$. If $\aa = 0$ then so is $\aa^\star$, and $\bK$ is already
simple without the $(\Phi, S)$ transformation being used. There
are therefore two canonical types for $\bK$,
\beq\label{E:2kv}
{\rm Type\, 1}:\; \bKa = V\partial_u, \quad {\rm or} \quad
{\rm Type\, 2}:\; \bKb = \partial_\zz + \partial_{\bar\zz}.
\eeq
These are asymptotically timelike and spacelike, respectively.

\section{Stationary solutions}

The obvious and easiest way to simplify the field equations was to assume that
the metric was stationary. The Type 2 Killing vectors are
asymptotically spacelike and so I assumed that $\EE$ had a Type 1
Killing vector $\bKa= V\partial_u$. The coordinates used in the
last section assumed that $P=1$. If we transform to the more
general coordinates where $P\neq 1$, using an A-transformation
(\ref{E:Atrans}) with associated change in the $(r,u)$ variables, we get
\beqn
\begin{split}
\bk^\prime = A\bk, \quad \bl^\prime =
A^{-1}\bl, r^\prime = A^{-1}r,\quad u^\prime = Au,\\
\bKa = V\partial_u = VA\partial_{u^\prime} =
\partial_{u^\prime}\quad {\textrm if}\; VA = 1.
\end{split}
\eeqn
The metric can therefore be assumed independent of $u$, but $P$
may not be constant. The basic functions, $L, P$ and $m$ are
functions of $(\zz,\bar\zz)$ alone, and the metric simplifies to
\beq\label{E:split1}
ds^2 = ds_o^2 + 2mr/(r^2 + \Sigma^2)k^2,
\eeq
where the ``base'' metric, is
\begin{subequations}\label{E:split2}
\begin{gather}
(ds_0)^2 = 2(r^2 + \Sigma^2)P^{-2}d\zz d\bar\zz -
2l_0k,\\
l_0 =dr + i(\Sigma_{,\zz} d \zz - \Sigma_{,\bar\zz} d \bar\zz) +
\left[ \half K - \frac{M\Sigma}{(r^2+\Sigma^2)}\right]k.
\end{gather}
\end{subequations}

Although the base metric is flat for Schwarzschild it is not
so in general. $\Sigma, K$ and $M$ are all functions of the
derivatives of $L$ and $P$,
\beq
\begin{split}\label{E:aa}
\Sigma &= P^2{\rm Im}(L_{\bar\zz}), \qquad
K = 2\nabla^2\ln P,\\
M &= \Sigma K +
\nabla^2\Sigma, \quad \nabla^2 = P^2\partial_\zz
\partial_{\bar\zz},
\end{split}
\eeq
The mass function, $m$,and $M$ are conjugate harmonic functions,
\beq\label{E:remainder2}
m_{\zz} = -iM_{\zz}, \quad
m_{\bar\zz} = +iM_{\bar\zz},
\eeq
and the remaining field equations are
\beq\label{E:remainder1}
\nabla^2 K = 0,\quad
\nabla^2 M = 0.
\eeq
If $m$ is a particular solution of these equations then so is $m +
m_0$ where $m_0$ is an arbitrary constant. The most general situation where
the metric splits in this way is when $P, L$ and $M$ are all independent of
u but $m = cu+A(\zz,\bar\zz)$. The field equations for these are given in
(\ref{E:withC}) and in Kerr (1963). We can state this as a theorem:
\begin{theorem}\label{TH:ks} If $ds_0^2$ is any stationary (diverging)
algebraically special metric, or more generally a solution of (\ref{E:withC}),
then so is
\beqn
ds_0^2 + \frac{2m_0 r}{r^2 + \Sigma^2}k^2,
\eeqn
where $m_0$ is an arbitrary constant. These are the most general diverging
algebraically special spaces that split in this way.
\end{theorem}
These are ``generalized Kerr--Schild'' metrics with base spaces
$ds_0^2$ that are not necessarily flat.

The field equations for stationary AS metrics are certainly
simpler than the original ones, (\ref{E:kfe}), but they are still
PDEs, not ODEs, and their complete solution is unknown.

\section{Axial symmetry}

We are getting close to Kerr. At this point I assumed that the
metric was axially symmetric as well as stationary. I should have
revisited the Killing equations to look for any Killing vector
(KV) that commutes with the stationary one, $\partial_u$. However,
I knew that it could not also be Type 1\footnote{If it were it would be
parallel to the stationary KV and therefore a constant multiple of it.}
and therefore it must be Type 2. It seemed fairly clear that it
could be transformed to the canonical form $i(\partial_\zz -
\partial_{\bar\zz})$ ($=
\partial_y$ where $\zz = x+iy$) or equivalently
$i(\zz\partial_\zz - \bar\zz\partial_{\bar\zz})$ ($=
\partial_\phi$ in polar coordinates where $\zz = Re^{i\phi}$).
I was getting quite eager at this point so I decided to just
assume such a KV and see what turned up\footnote{All possible
symmetry groups were found for diverging AS spaces in George
C. Debney's Ph.D.\ thesis. My 1963 expectations were confirmed there.}.

From the first equation in (\ref{E:remainder1}), the curvature
$\nabla^2(\ln P)$ of the 2-metric $P^{-2}d\zz d\bar\zz$ is a
harmonic function,
\beqn
\nabla^2\ln P = P^2(\ln P)_{,\zz\bar\zz} = F(\zz) +
\bar F(\bar\zz),
\eeqn
where $F$ is analytic.. There is only one known
solution of this equation for $F$ not a constant,
\beq\label{}
P^2 = P_0(\zz+\bar\zz)^3, \quad \nabla^2\ln P =
-\frac{3}{2}P_0(\zz+\bar\zz),
\eeq
where $P_0$ is an arbitrary constant. The mass function $m$ is
then constant and the last field equation, $\nabla^2M = 0$, can be
solved for $L$. The final metric is given in Kerr and Debney
(1969), equation (6.14), but it is not worth writing out here
since it is not asymptotically flat.

If $\EE$ is to be the metric for a localized physical source then
the null congruence should be asymptotically the same as
Schwarzschild. $F(\zz)$ must be regular everywhere, including at
infinity, and must therefore be constant,
\beq\label{}
\half K = PP_{,\zz\bar\zz}-P_{,\zz}P_{,\bar\zz} = R_0 = \pm P_0^2,
\quad {\rm (say)}.
\eeq

As was shown in Kerr and Debney (1969), the appropriate
Killing equations for a $\bKb$ that commutes with $\bKa$ are
\beq\label{E:keq}
\begin{split}
\bKb &= \aa\dz + \bar\aa\dzb,\quad \aa =\aa(\zz),\\
\bKb L &= - \aa_\zz L,\qquad \,\,\bKb \Sigma = 0,\\
\bKb P &= {\rm Re}(\aa_\zz) P,\quad \bKb m =
0.
\end{split}
\eeq

I do not remember the choice I made for the canonical form for
$\bKb$ in 1963, but it was probably $\partial_y$. The choice in
Kerr and Debney (1969) was
\beqn
\aa = i\zz,\quad\Rightarrow\quad \bKb = i(\zz\partial_\zz -
\bar\zz\partial_{\bar\zz}),
\eeqn
and that will be assumed here.
For any function $f(\zz,\bar\zz)$,
$$ \bKb f = 0 \quad\Rightarrow\quad f(\zz,\bar\zz) = g(Z),
\quad {\rm where}\quad Z =\zz\bar\zz.
$$
Now ${\rm Re}(\aa_{,\zz}) = 0$, and therefore
$$
\bKb P = 0, \quad \Rightarrow \quad P= P(Z),
$$
and therefore $K$ is given by
\beqn
\half K =P^2(\ln P)_{,\zz\bar\zz} = PP_{,\zz\bar\zz}-P_{,\zz}P_{,\bar\zz}
=Z_0\quad \Rightarrow \quad P = Z + Z_0,
\eeqn
after a $\Phi(\zz)$-coordinate transformation.
Note that the form of the metric is invariant under the transformation
\beq\label{E:trans}
\begin{split}
r = A_0 &r^*,\quad u = A_0^{-1}u^*,\quad \zz = A_0\zz^*,\\
Z_0 &= A_0^{-2}Z_0^*, \quad m_0 = A_0^{-3}m_0^*,
\end{split}
\eeq
where $A_0$ is a constant, and therefore $Z_0$ is a disposable constant.
We will choose it later.

The general solution of (\ref{E:keq}) for $L$ and $\Sigma$ is
\beqn
L = i\bar\zz P^{-2}B(Z), \quad \Sigma = ZB^\prime - (1-Z_0
P^{-1})B,
\eeqn
where $B^\prime = dB/dZ$. The complex ``mass'', $m+iM$, is an
analytic function of $\zz$ from (\ref{E:remainder2}), and is also a
function of $Z$ from (\ref{E:keq}). It must therefore be a constant,
\beqn
m+iM = \mu_0 = m_0 + iM_0.
\eeqn

Substituting this into (\ref{E:aa}), the equation for $\Sigma$,
\beqn
\begin{split}
\Sigma K + \nabla^2\Sigma = M &= M_0\quad\longrightarrow\\
P^2[Z\Sigma^{\prime\prime} + \Sigma^\prime] + 2&Z_0\Sigma = M_0.
\end{split}
\eeqn
The complete solution to this is
\beqn
\Sigma = C_0 + \frac{Z-Z_0}{Z+Z_0}[-a + C_2\ln Z],\\
\eeqn
where $\{C_0, a, C_2\}$ are arbitrary constants. This gave a four--parameter
metric when these known functions are substituted into (\ref{E:split1})
and (\ref{E:split2}). However, if $C_2$ is nonzero then the final metric
is singular at $r=0$. It was therefore omitted in Kerr (1963). The
``imaginary mass'' is then $M = 2Z_0C_0$ and so $C_0$ is a multiple of
the NUT parameter. It was known in 1963 that the metric cannot be
asymptotically flat if this is nonzero and so it was also omitted. The
only constants retained were $m_0, a$ and $Z_0$. When $a$ is zero and
$Z_0$ is positive the metric is that of Schwarzschild. It was not clear
that the metric would be physically interesting when $a \neq 0$, but if
it had not been so then this whole exercise would have been futile.

The curvature of the 2-metric, $2P^{-2}d\zz d\bar\zz$, had to have the
same sign as Schwarzschild if the metric was to be asymptotically
flat, and so $Z_0 = +P_0^2$. The basic functions in the metric then became
\begin{gather*}
Z_0 = P_0^2,\quad P = \zz\bar\zz + P_0^2, \quad m = m_0,\quad M = 0,\\
L = ia\bar\zz P^{-2},\quad \Sigma = -a\frac{\zz\bar\zz-Z_0}{\zz\bar\zz
+Z_0}.
\end{gather*}

The metric was originally published in spherical polar coordinates. The
relationship between these and the $(\zz, \bar\zz)$ coordinates is
\beqn
\zz = P_0 {\rm cot} \tfrac{\theta}{2} e^{i\phi}.
\eeqn
At this point we choose $A_0$ in the transformation (\ref{E:trans}) so that
\beqn
2P_0^2 = 1, \quad \Rightarrow \quad k = du + a \,{\rm sin}^2\theta d\phi
\eeqn
Recalling the split of (\ref{E:split1}) and (\ref{E:split2}),
\beq\label{E:Ka}
ds^2 = ds_0^2 + 2mr/(r^2 + a^2 {\rm cos}^2\theta)k^2\\
\eeq
where $m=m_0$, a constant, and
\beq\label{E:Kb}
\begin{split}
ds_0^2 = &(r^2 + a^2{\rm cos}^2\theta)(d\theta^2+{\rm sin}^2\theta d\phi^2)\\
-&(2dr +du - a\,{\rm sin}^2\theta d\phi)(du + a \,{\rm sin}^2\theta d\phi).
\end{split}
\eeq
This is the original form of Kerr (1963), except that $u$ has been
replaced by $-u$ to agree with current conventions, and $a$ has been
replaced with its negative\footnote{We will see why later.}.

Having found this fairly simple metric, I was desperate to see whether
it was rotating. Fortunately, I knew that the curvature of the base
metric, $ds_0^2$, was zero, and so it was only necessary to find
coordinates where this was manifestly Minkowskian. These were
\beqn
(r+ia)e^{i\phi}{\rm sin}\theta = x+iy,\quad r\,{\rm cos}\theta = z,
\quad r+u=-t.
\eeqn
This gives the Kerr--Schild form of the metric,
\beq\label{E:KSform}
\begin{split}
ds^2 = d&x^2 + dy^2 + dz^2 - dt^2 + \frac{2mr^3}{r^4+ a^2z^2}[dt +
\frac{z}{r}dz\\
&+ \frac{r}{r^2+a^2}(xdx+ydy) - \frac{a}{r^2+a^2}(xdy-ydx)]^2.
\end{split}
\eeq
where the surfaces of constant $r$ are confocal ellipsoids of
revolution about the z-axis,
\beq\label{E:r}
\frac{x^2+y^2}{r^2+a^2} + \frac{z^2}{r^2} = 1.
\eeq
Asymptotically, of course, $r$ is just the distance from the origin
in the Minkowskian coordinates, and the metric is clearly asymptotically flat.

\subsection*{Angular momentum}

After the metric had been put into its Kerr--Schild form I went to
Alfred Schild and told him I was about to calculate the angular
momentum of the central body. He was just as excited as I was and so he
joined me in my office while I computed. We were both heavy smokers at
that time, so you can imagine what the atmosphere was like, Alfred
puffing away at his pipe in an old arm chair, and myself chain--smoking
cigarettes at my desk.

The Kerr--Schild form is particularly suitable for calculating the
physical parameters of the solution. My PhD thesis at Cambridge was
entitled ``Equations of Motion in General Relativity''. It had been
claimed previously in the literature that it was only necessary to
satisfy the momentum equations for singular particles to be able to
integrate the EIH quasi-static approximation equations at each order.
One thing shown in my thesis was the physically obvious fact that the
angular momentum equations were equally important. Some of this was
published in Kerr (1958) and (1960). Because of this previous work
I was well aware how to calculate the angular momentum in this new metric.

It was first expanded in powers of $R^{-1}$, where $R=x^2+y^2+z^2$ is
the usual Euclidean distance,
\beq\label{E:approx1}
\begin{split}
ds^2 = &dx^2+dy^2+dz^2-dt^2 + \frac{2m}{R}(dt+dR)^2\\
&-\frac{4ma}{R^3}(xdy-ydx)(dt+dR) + O(R^{-3})
\end{split}
\eeq
Now, if $x^\mu \rightarrow x^\mu +a^\mu$ is an infinitesimal coordinate
transformation, then $ds^2 \rightarrow ds^2 + 2 da_\mu dx^\mu$. If we choose
\beqn
\begin{split}
a_\mu dx^\mu &= -\frac{am}{R^2}( xdy-ydx)\quad \Rightarrow\\
2da_\mu dx^\mu &= -4m\frac{4am}{R^3}(x dy - y dx)dR,
\end{split}
\eeqn
then the approximation in (\ref{E:approx1}) simplifies to
\beq\label{E:approx2}
\begin{split}
ds^2 = &dx^2+dy^2+dz^2-dt^2 + \frac{2m}{R}(dt+dR)^2\\
&-\frac{4ma}{R^3}(xdy-ydx)dt + O(R^{-3}).
\end{split}
\eeq
The leading terms in the linear approximation for the gravitational field
around a rotating body were well known (for instance, see Papapetrou (1974)
or Kerr (1960)). The contribution from the angular momentum vector,
$\bf J$, is
$$
4R^{-3}\ee_{ijk}J^i x^j dx^k dt.
$$
A comparison of the last two equations showed that the physical
parameters
were
$$
{\rm Mass} = m, \quad {\bf J} = (0,0,ma).\footnote{Unfortunately, I was
rather hurried when performing this calculation and got the sign wrong.
This is why the sign of the parameter $a$ in Kerr(1963) is different to
that in all other publications, including this one. This way of
calculating $\bf J$ was explained at the First Texas Symposium at the
end of 1963, see Kerr(1965), but I did not check the sign at that time.}
$$
When I turned to Alfred Schild, who was still sitting in the arm--chair
smoking away, and said ``Its rotating!'' he was even more excited than
I was. I do not remember how we celebrated, but celebrate we did!

Robert Boyer subsequently calculated the angular momentum by comparing
the known Lense--Thirring results for frame dragging around a rotating
object in linearized relativity with the frame dragging for a circular
orbit in a Kerr metric. This was a very obtuse way of calculating the
angular momentum since the approximation (\ref{E:approx2}) was the basis
for the calculations by Lense and Thirring, but it did show that the
sign was wrong in the original paper!

\section{Singularities and Topology}

The first Texas Symposium on Relativistic Astrophysics was held in
Dallas December 16-18, 1963, just a few months after the discovery of
the rotating solution. It was organized by a combined group of
Relativists and Astrophysicists and its purpose was to try to find an
explanation for the newly discovered quasars. The source 3C273B had been
observed in March and was thought to be about a million million times
brighter than the sun.

It had been long known that a spherically symmetric body could collapse
inside an event horizon to become what was to be later called a black
hole by John Wheeler. However, the Schwarzschild solution was non--rotating
and it was not known what would happen if rotation was present. I presented
a paper called ``Gravitational collapse and rotation'' in which I outlined
the Kerr solution and said that the topological and physical properties
of the event horizon may change radically when rotation is taken into
account. It was not known at that time that Kerr was the only possible
stationary solution for such a rotating black hole and so I discussed
it as an example of such an object.

Although this was not pointed out in the original letter, Kerr (1963),
the geometry of Kerr is even more complicated than the Kruskal extension
of Schwarzschild. The metric is everywhere nonsingular, except on the ring
$$
z = 0,\qquad x^2 + y^2 = a^2.
$$
The Weyl scalar, $R_{abcd}R^{abcd} \rightarrow \infty$ near these points
and so they are true singularities, not just coordinate ones. Furthermore,
this ring behaves like a branch point in the complex plane. If one travels
on a closed curve that threads the ring the initial and final metrics are
different: $r$ changes sign. Equation (\ref{E:r}) has one nonnegative root
for $r^2$, and therefore two real roots, $r_\pm$, for $r$. These coincide
where $r^2 =0$, i.e., on the disc $D$ bounded by the ring singularity
$$
\hbox{$D$:}\quad z = 0, \quad x^2+y^2 \leq a^2.
$$
The disc can be taken as a branch cut for the analytic function $r$. We
have to take two spaces, $E_1$ and $E_2$ with the topology of $R^4$ less
the disc $D$. The points above $D$ in $E_1$ are joined to the points below
$D$ in $E_2$ and vice versa. In $E_1$, $r>0$ and the mass is positive at
infinity; in $E_2$, $r<0$ and the mass is negative. The metric is then
everywhere analytic except on the ring.

It was trivially obvious to everyone that if the parameter $a$ is very
much less than $m$ then the Schwarzschild event horizon at $r=2m$ will
be modified slightly but cannot disappear. For instance, the light cones
at $r=m$ in Kerr all point inwards for small $a$. Before I went to the
meeting I had calculated the behaviour of the time like geodesics up and
down the axis of rotation and found that horizons occurred at the points
on the axis in $E_1$ where
$$
r^2 -2mr+ a^2 = 0, \quad r = |z|.
$$
but that there are no horizons in $E_2$ where the mass is negative. In
effect, the ring singularity is ``naked'' in that sheet.

I made a rather hurried calculation of the two event horizons in $E_1$
before I went to the Dallas Symposium and claimed incorrectly there, Kerr
(1964), that the equations for them were the two roots of
$$
r^4 - 2mr^3 + a^2z^2 = 0,
$$
whereas $z^2$ should be replaced by $r^2$ in this and the true equation is
$$
r^2 - 2mr +a^2 = 0.
$$
I attempted to calculate this using inappropriate coordinates and
assuming that the equation would be: ``$\psi(r,z)$ is null for some
function of both $r$ and $z$''. I did not realize that this function
depended only on $r$.

The Kerr--Schild coordinates are a generalization of the
Eddington--Finkelstein coordinates for Schwarzschild. For the latter
future--pointing radial geodesics are well behaved but not those traveling
to the past. Kruskal coordinates were designed to handle both. Similarly
for Kerr, the coordinates given here only handle ingoing curves. This
metric is known to be Type D and therefore it has another set of
Debever--Penrose vectors and an associated coordinate system for which
the outgoing geodesics are well behaved, but not the ingoing ones.

This metric consists of three blocks, outside the outer event horizon,
between the two horizons and within the inner horizon (at least for
$m < a$, which is probably true for all existing black holes). Just as
Kruskal extends Schwarzschild by adding extra blocks, Boyer and Lindquist
(1967) and Carter (1968) independently showed that the maximal extension
of Kerr has a similar proliferation of blocks. However, the Kruskal
extension has no application to a real black hole formed by the collapse
of a spherically symmetric body and the same is true for Kerr. In fact,
even what I call $E_2$, the sheet where the mass is negative, is probably
irrelevant for the final state of a collapsing rotating object.

Ever since this metric was first discovered people have tried to fit an
interior solution. One morning during the summer of 1964 Ray Sachs and
myself decided that we would try to do so. Since the original form is
useless and the Kerr--Schild form was clearly inappropriate we started
by transforming to the canonical coordinates for stationary axisymmetric
solutions.

In Papapetrou (1966) there is a very elegant treatment of stationary
axisymmetric Einstein spaces. He shows that if there is a real
non-singular axis of rotation then the coordinates can be chosen so that
there is only one off-diagonal component of the metric. We call such a
metric quasi-diagonalize. All cross terms between $\{dr,d\theta\}$ and
$\{dt,d\phi\}$ can be eliminated by transformations of the type
$$dt^\prime = dt + Adr + Bd\theta, \quad d\phi^\prime
= d\phi + Cdr + Dd\theta.$$
where the coefficients can be found algebraically. Papapetrou proved that
$dt^\prime$ and $d\phi^\prime$ are perfect differentials if the axis is
regular\footnote{It is shown in Kerr and Weir (1976) that if the metric
is also algebraically special then it is quasi-diagonalize precisely when
it is Type D. These metrics are the NUT parameter generalization of Kerr.}.

Ray and I calculated the coefficients $A\dots D$ and transformed the metric
to the Boyer-Lindquist form,
\begin{align*}\label{}
dt &\rightarrow dt + \frac{2mr}{\Delta}dr\\
d\phi &\rightarrow -d\phi + \frac{a}{\Delta}dr\\
\Delta &= r^2 - 2m r + a^2,\\
\Sigma &= r^2 + a^2 {\rm cos}^2\theta,
\end{align*}
where, as before, $u = -(t + r)$. The right hand sides of the first two
equations are clearly perfect differentials as the Papapetrou analysis
showed. The full Boyer-Lindquist form of the metric is
\beq\label{E:BL}
\begin{split}
ds^2 = &\frac{\Sigma}{\Delta}dr^2 - \frac{\Delta}{\Sigma}
[dt - a {\rm sin}^2\theta d\phi]^2 +\\
&\Sigma d\theta^2 + \frac{{\rm sin}^2\theta}{\Sigma}[(r^2+a^2)d\phi -a dt]^2,
\end{split}
\eeq
after some tedious analysis that used to be easy but now seems to require
an algebraic package such as Maple.

Having derived this canonical form, we studied the metric for at least
ten minutes and then decided that we had no idea how to introduce a
reasonable source into a metric of this form, and probably would never
have. Presumably those who have tried to solve this problem in the last
43 years have had similar reactions. Soon after this failed attempt
Robert Boyer came to Austin. He said to me that he had found a new
quasi--diagonalized form of the metric. I said ``Yes. It is the one with
the polynomial $r^2 - 2mr +a^2$'' but for some reason he refused to believe
that we had also found this form. Since it did not seem a ``big deal'' at
that time I did not pursue the matter further, but our relations were
hardly cordial after that.

One of the main advantages of this form is that the event horizons can
be easily calculated since the inverse metric is simple. If
$f(r, \theta) = 0$ is a null surface then
$$
\Delta(r) f_{,rr} + f_{,\theta\theta} = 0,
$$
and therefore $\Delta \leq 0$. The two event horizons are the surfaces
$r = r_\pm$ where the parameters $r_\pm$ are the roots of $\Delta = 0$,
$$
\Delta = r^2 - 2mr + a^2 = (r - r_+)(r - r_-).
$$

If $a < m$ there are two distinct horizons between which all time-like
lines point inwards; if $a= m$ there is only one event horizon; and
for larger $a$ the singularity is bare! Presumably, any collapsing star
can only form a black hole if the angular momentum is small enough:
$a < m$. This seems to be saying that the body cannot rotate faster
than light, if the final picture is that the mass is located on the
ring radius $a$. However, it should be remembered that
this radius is purely a coordinate radius, and that there is no way that
the final stage of such a collapse is that all the mass is located at
the singularity.

The reason for the last statement is that if the mass where to end on
the ring then there would be no way to avoid the second asymptotically
flat sheet where the mass appears negative. I do not believe that the
star opens up like this along the axis of rotation. If we remember that
the metric is discontinuous across the disc bounded by the singular ring
then it is quite possible that a well-behaved finite source could be put
between $z=0_\pm, |R|< a$\footnote{This has been done using
$\delta$--functions, but I am thinking more of a nonsingular source where
the distance between the two sides of the disk is nonzero.}. This would
correspond to the surface of the final body being $r=0$ in Boyer-Lindquist
coordinates, say, but where the interior corresponds to $r<0$.. The actual
surface may be more complicated than this but I am quite sure that this is
closer to the final situation than that the matter all collapses onto the
ring.

What I believe to be more likely is that the inner event horizon never
actually forms. As the body continues to collapse inside its event horizon
it spins faster and faster so that the geometry in the region between its
outer surface and the outer event horizon approaches that between the two
event horizons for Kerr. The surface of the body surface will appear to
be asymptotically null. The full metric may not be geodesically complete.
Many theorems have been claimed stating that a singularity must exist if
certain conditions are satisfied, but they all make assumptions that may
not be true for collapse to a black hole. Furthermore, these assumptions
are often (usually?) unstated or unrecognised, and the proofs are
dependent on other claims/theorems that may not be correct.

These are only two of a very large range of possibilities for the interior.
What happens after the outer horizon forms is still a mystery after more
than four decades. It is also the main reason why I said at the end of
Kerr (1963) that ``It would be desirable to calculate an interior
solution\dots''. This statement has been taken by some to mean that
I thought the metric only represented a real rotating star. This is untrue
and is an insult to all those relativists of that era who had been looking
for such a metric to see whether the event horizon of Schwarzschild would
generalise to rotating singularities.

The metric was known to be Type D with two distinct geodesic and shearfree
congruences from the moment it was discovered. This means that if the other
is used instead of $k$ then the metric must have the same form, i.e., it
is invariant under a finite transformation that reverses ``time'' and
possibly the axis of rotation in the appropriate coordinates. This
also meant that there is an extension that is similar to the
Kruskal--Szekeres extension of Schwarzschild. Both Boyer and Lindquist
(1967) and a fellow Australasian, Brandon Carter (1968), solved the
problem of constructing the maximal extensions of Kerr, and even that
for charged Kerr. These are mathematically fascinating and the latter
paper is a beautiful analysis of the problem, but the final result is
of limited physical significance.

Brandon Carter's (1968) paper was one of the most significant papers on
the Kerr metric during the mid-sixties for another reason. He showed that
there is an extra invariant for geodesic motion which is quadratic in the
momentum components: $J =X_{ab}v^av^b$ where $X_{ab}$ is a Killing tensor,
$X_{(ab;c)} = 0$. This gave a total of four invariants with the two Killing
vector invariants and $|{\bf v}|^2$ itself, enough to generate a complete
first integral of the geodesic equations.

Another significant development was the ``proof'' that this is the
only stationary metric with a simply connected bounded event horizon, i.e.,
the only possible black hole. Many contributed to this, including
Steven Hawking (1972), Brandon Carter (1971) and another New Zealander,
David Robinson (1975). The subsequent work in this area is discussed by
David in an excellent article in this book and so I will not pursue
this any further here.

\section{Kerr--Schild metrics}
One morning during the fall semester 1963, sometime before the First Texas
Symposium, I tried generalizing the way that the field equations split for
the Kerr metric by setting
$$\label{E:gKS}
ds^2 = ds^2_0 + \frac{2mr}{r^2 + \Sigma^2} k^2.
$$
The base metric $ds_0^2$ was to be an algebraically special metric with
$m=0$. From an initial rough calculation this had to be flat. Also, it
seemed that the coordinates could be manipulated so that
$\partial L = L_\zz - L L_u= 0$ and that the final metric depended on an
arbitrary analytic function of the complex variable $\zz$. At this point
I lost interest since the metric had to be singular at the poles of the
analytic function unless this function was quadratic and the metric was
therefore Kerr.

Sometime after the Texas Symposium Jerzy Plebanski visited Austin. Alfred
Schild gave one of his excellent parties for Jerzy during which I heard
them talking about solutions of the Kerr--Schild type\footnote{This name
came later, of course.},\break $ds^2_0 + h k^2$, where the first term is flat
and $k$ is any null vector. I commented that there might be some
algebraically special spaces with this structure depending on an arbitrary
function of a complex variable but that this had not been checked.

At this point Alfred and I retired to his home office and calculated the
simplest field equation, $R_{ab}k^a k^b = 0$. To our surprise this showed
that the null vector had to be geodesic. We then calculated
$k_{[a} R_{b]pq[c}k_{d]} k^p k^q$, found it to be zero and deduced that
all metrics of this type had to be algebraically special. This meant that
all such spaces with a diverging congruence might already be known. We
checked my original calculations next day and found them to be correct.

As was stated in Theorem~\ref{TH:ks}, $m$ is a unique function of $P$ and
$L$ unless there is a canonical coordinate system where $m$ is linear in
$u$ and $\{L,P\}$ are functions of $\{\zz,\bar\zz\}$ alone. If the base
space is flat then $m_{,u}=c = 0$ and the metric is stationary. The way
these Kerr--Schild metrics were found originally was by showing that in a
coordinate system where $P =1$ the canonical coordinates could be chosen
so that $\partial L = 0$. Transforming from these coordinates to ones
where $P \neq 1$ and $\partial_u$ is a Killing vector,
\beq\label{E:ksL}
P_{,\zz\zz} = 0, \quad L = P^{-2} \bar\phi(\bar\zz),
\eeq
where $\phi(\zz)$ is analytic. From the first of these $P$ is a real
bilinear function of $\zz$ and therefore of $\bar \zz$,
$$
P = p\zz\bar\zz + q\zz + \bar q \bar\zz + c.
$$
This can be simplified to one of three canonical forms,
$P = 1, 1\pm \zz\bar \zz$ by a linear transformation on $\zz$.
We will assume henceforth that
\beqn
P = 1 + \zz\bar\zz.
\eeqn
The only problem was that this analysis depended on results for
algebraically special metrics and these had not been published and would
not be for several years. We had to derive the same results by a more
direct method. The metric was written as
\begin{subequations}\label{E:ks}
\begin{align}
ds^2 &= dx^2+dy^2+dz^2-dt^2 + h k^2,\\
k &= (du+\bar Y d\zeta + Y d\bar\zz + Y\bar Y dv)/(1+Y\bar Y),
\end{align}
\end{subequations}
where $Y$ is the old coordinate $\zz$ used in (\ref{E:ksL})\footnote{$Y$
is the ratio of the two components of the spinor corresponding to $\bk$.} and
$$ u = z+t,\quad v = t-z, \quad \zz = x+iy\footnote{Note that there certain
factors of $\sqrt{2}$ have been omitted to simplify the results. This does
lead to a factor 2 appearing in (\ref{E:hh}).}.$$

The tetrad used to calculate the field equations was defined naturally from
the identity
\begin{equation*}
ds^2 = (d\zz + Ydv)(d\bar\zz+\bar Ydv)-(dv+hk)k
\end{equation*}

Each of these spaces has a symmetry which is also a translational symmetry
for the base Minkowski space, $ds^2_0$. The most interesting situation is
when this is time-like and so it will be assumed that the metric is
independent of $t = \half(u+v)$.

If $\phi(Y)$ is the same analytic function as in (\ref{E:ksL}) then $Y$
is determined as a function of the coordinates by
\beq\label{E:Y}
Y^2\bar\zz +2z Y- \zz + \phi(Y) = 0.
\eeq
and the coefficient of $k^2$ in (\ref{E:ks}) is
\beq\label{E:hh}
h = 2m {\rm Re} (2Y_\zz),
\eeq
where $m$ is a real constant.
Differentiating (\ref{E:Y}) with respect to $\zz$ gives
\beq\label{E:ksh}
Y_\zz = (2Y\bar\zz + 2z + \phi^\prime )^{-1}.
\eeq
The Weyl spinor invariant is given by
$$
\Psi_2 = c_0 m Y^3_\zz,
$$
where $c_0$ is some power of 2, and the metric is therefore singular
precisely where Y is a repeated root of its defining equation, (\ref{E:Y}).

If the k-lines are projected onto the Euclidean 3-space $t=0$ with
$\{x,y,z\}$ as coordinates so that $ds_E^2 = dx^2+dy^2+dz^2$, then the
perpendicular from the origin meets the projected k-line at the point
\beqn
F_0:\quad \zz = \frac{\phi-Y^2\bar\phi}{P^2},\quad z = -\frac{\bar Y\phi
+ Y\bar\phi}{P^2},
\eeqn
and the distance of the line from the origin is
$$ D = \frac{|\phi|}{1+Y\bar Y},$$
a remarkably simple result. This was used by Kerr and Wilson (1978) to
prove that unless $\phi$ is quadratic the singularities are unbounded and
the spaces are not asymptotically flat. The reason why I did not initially
take the general Kerr--Schild metric seriously was that this was what
I expected.

Another point that is easily calculated is $Z_0$ where the line meets the
plane $z=0$,
\beqn
Z_0:\quad \zz = \frac{\phi+Y^2\bar\phi}{1-(Y\bar Y)^2},\quad z = 0,
\eeqn

The original metric of this type is Kerr where
$$\label{E:kerrphi}
\phi(Y) = -2iaY,\quad D= \frac{2|a||Y|}{1+|Y|^2} \leq |a|.
$$
If $\phi(Y)$ is any other quadratic function then it can be transformed to
the same value by using an appropriate Euclidean rotation and translation
about the t-axis. The points $F_0$ and $Z_0$ are the same for Kerr, so
that $F_0$ lies in the z-plane and the line cuts this plane at a point
inside the singular ring provided $|Y| \neq 1$. The lines where $|Y| = 1$
are the tangents to the singular ring lying entirely in the plane $z = 0$
outside the ring. When $a \rightarrow 0$ the metric becomes Schwarzschild
and all the $Y$-lines pass through the origin.

When $\phi(Y) = -2iaY$ (\ref{E:Y}) becomes
$$
Y^2 \bar\zz + 2(z-ia)Y -\zz = 0.
$$
There are two roots, $Y_1$ and $Y_2$ of this equation,
\begin{gather*}
\begin{split}
Y_1 &= \frac{r\zz}{(z+r)(r-ia)},\quad 2Y_{1,\zz} =
+\frac{r^3+iarz}{r^4+a^2z^2}\\
Y_2 &= \frac{r\zz}{(z-r)(r+ia)},\quad 2Y_{2,\zz} =
-\frac{r^3+iarz}{r^4+a^2z^2}
\end{split}
\end{gather*}
where r is a real root of (\ref{E:r}). This is a quadratic equation for
$r^2$ with only one nonnegative root and therefore two real roots differing
only by sign, $\pm r$. When these are interchanged, $r \leftrightarrow -r$,
the corresponding values for $Y$ are also swapped, $Y_1 \leftrightarrow Y_2$.

When $Y_2$ is substituted into the metric then the same solution is returned
except that the mass has changed sign. This is the other sheet where $r$
has become negative. It is usually assumed that $Y$ is the first of these
roots, $Y_1$. The coefficient $h$ of $k^2$ in the metric, (\ref{E:ks}), is then
$$
h = 2m {\rm Re} (2Y_\zz)=\frac{2mr^3}{r^4 + a^2z^2}.
$$
This gives the metric in its KS form,(\ref{E:KSform}).

The results were published in two places, Kerr and Schild (1965a,b). The
first of these was a talk that Alfred gave at the Galileo Centennial in
Italy, the second was an invited talk that I gave, but Alfred wrote, at the
Symposium on Applied Mathematics of the American Mathematical Society,
April 25 1964. The manuscript had to be presented before the conference
so that the participants had some chance of understanding results from
distant fields. We stated on page 205 that
{\quote
``Together with their graduate student, Mr. George Debney, the authors
have examined solutions of the nonvacuum Einstein-Maxwell equations where
the metric has the form (2.1)\footnote{Equation (\ref{E:ks}) in this paper.
It refers to the usual Kerr--Schild ansatz.}. Most of the results mentioned
above apply to this more general case. This work is continuing.''}.

\section{Charged Kerr}

What was this quote referring to? When we had finished with the Kerr--Schild
metrics, we looked at the same problem with a nonzero electromagnetic field.
The first stumbling block was that $R_{ab}k^ak^b=0$ no longer implied that
the k-lines are geodesic. The equations were quite intractable without this
and so it had to be added as an additional assumption. It then followed that
the principal null vectors were shearfree, so that the metrics had to be
algebraically special. The general forms of the gravitational and
electromagnetic fields were calculated from the easier field equations. The
E--M field proved to depend on two functions called $A$ and $\gamma$ in
Debney, Kerr and Schild (1969).

When $\gamma = 0$ the ``difficult'' equations are linear and similar to
those for the purely gravitational case. They were readily solved giving
a charged generalization of the original Kerr--Schild metrics. The
congruences are the same as for the uncharged metrics, but the coefficient
of $k^2$ is
\beq\label{E:chargedh}
h = 2m {\rm Re}(2Y_{,\zz}) - |\psi|^2|2Y_{,\zz}|^2.
\eeq
where $\psi(Y)$ is an extra analytic function generating the
electromagnetic field. This is best expressed through a potential,
\begin{align*}
f &= \half F_{\mu\nu}dx^\mu dx^\nu = -d\aa,\\
\aa &=-P(\psi Z +\bar\psi\bar Z)k -\half(\chi d\bar Y + \bar\chi dY),
\end{align*}
where
\beqn
\chi = \int P^{-2} \psi(Y) dY,
\eeqn
$\bar Y$ being kept constant in this integration.

The most important member of this class is charged Kerr. For this,
\beq\label{}
h = \frac{2mr^3- |\psi(Y)|^2r^2}{r^4 + a^2z^2}.
\eeq
Asymptotically, $r = R$, $k =dt-dR$, a radial null-vector and
$Y= {\rm tan}(\half \theta) e^{i\phi}$. If the analytic function $\psi(Y)$
is nonconstant then it must be singular somewhere on the unit sphere and so
the gravitational and electromagnetic fields will be also. The only
physically significant charged Kerr--Schild is therefore when $\psi$ is a
complex constant, $e + ib$. The imaginary part, $b$, can be ignored as it
gives a magnetic monopole, and so we are left with $\psi = e$, the electric
charge,
\beq\label{}
\begin{split}
ds^2 = d&x^2 + dy^2 + dz^2 - dt^2 + \frac{2mr^3-e^2r^2}{r^4+ a^2z^2}
[dt +\frac{z}{r}dz\\
&+ \frac{r}{r^2+a^2}(xdx+ydy) - \frac{a}{r^2+a^2}(xdx-ydy)]^2,
\end{split}
\eeq
The electromagnetic potential is
$$
\aa = \frac{er^3}{r^4+a^2z^2}\left[dt - \frac{a(xdy-ydx)}{r^2+a^2}\right],
$$
where a pure gradient has been dropped. The electromagnetic field is
\beqn
\begin{split}
(F_{xt}-iF_{yz}&, F_{yt}-iF_{zx}, F_{zt}-iF_{xy})\\
& =\frac{er^3}{(r^2+iaz)^3}(x,y,z+ia).
\end{split}
\eeqn
In the asymptotic region this field reduces to an electric field,
$$
{\bf E} = \frac{e}{R^3}(x,y,z),
$$
and a magnetic field,
$$
{\bf H} = \frac{ea}{R^5}(3xz, 3yz, 3z^2 - R^2).
$$
This is the electromagnetic field of a body with charge $e$ and magnetic
moment $(0,0,ea)$. The gyromagnetic ratio is therefore $ma/ea = m/e$, the
same as that for the Dirac electron. This was first noticed by
Brandon Carter and was something that fascinated Alfred Schild.

This was the stage we had got to before March 1964. We were unable to
solve the equations where the function $\gamma$ was nonzero so we enlisted
the help of our graduate student, George Debney. Eventually we realized that
we were unable to solve the more general equations and so we suggested to
George that he drop this investigation. He then tackled the problem of
finding all possible groups of symmetries in diverging algebraically special
spaces. He succeeded very well with this, solving many of the ensuing field
equations for the associated metrics. This work formed the basis for his
PhD thesis and was eventually published in Kerr and Debney (1970).

In Janis and Newman (1965) and Newman and Janis (1965) the authors
defined and calculated multipole moments for the Kerr metric, using the
Kerr--Schild coordinates as given in Kerr (1963). They then claimed that
this metric is that of a ring of mass rotating about its axis of symmetry.
Unfortunately, this cannot be so because the metric is multivalued on its
symmetry axis and is consequently discontinuous there. The only way that
this can be avoided is by assuming that the space contains matter on the
axis near the centre. As was acknowledged in a footnote to the second
paper, this was pointed out to the authors by the referee and myself before
the paper was published, but they still persisted with their claim.

\section{Newman's construction of the Kerr--Newman metric}

Newman knew that the Schwarzschild, Reissner--Nordstr\"{o}m and Kerr
metrics all have the same simple form in Eddington--Finkelstein or
Kerr--Schild coordinates (see eq.\ \ref{E:ks}). Schwarzschild and its
charged generalisation have the same null congruence, $k=dr-dt$; only
the coefficient $h$ is different,
$$
h=\frac{2m}{r} \quad \longrightarrow \quad h=\frac{2m}{r}-\frac{e^2}{r^2}.
$$
For these the complex divergence of the underlying null congruence is
$$\rho = \bar\rho= 1/r.
$$

Newman hoped to find a charged metric with the same congruence as Kerr but
with $h$ generalised to something like the Reissner--Nordstr\"{o}m form
with $e^2/r^2$ replaced by $e^2\rho^2$. This does not quite work since
$\rho$ is complex for Kerr so he had to replace $\rho^2$ with something real.

There are many real rational functions of $\rho$ and $\bar\rho$ that reduce
to $\rho^2$ when $\rho$ is real, so he wrote down several possibilities and
distributed them to his graduate students. Each was checked to see whether
it was a solution of the Einstein--Maxwell equations. The simplest,
$\rho^2\rightarrow\rho\bar\rho$, worked! The appropriate electromagnetic
field was then calculated, a non-trivial problem.

The reason that this approach was successful has nothing to do with
``complexifying the Schwarzschild and Reissner--Nordstr\"{o}m metrics'' by
some complex coordinate transformation, as stated in the original papers.
It works because all these metrics are of Kerr--Schild form and the
{\it general} Kerr--Schild metric can be charged by replacing the uncharged
$h$ with its appropriate charged version,
$h = 2m {\rm Re} (2Y_\zz) - |\psi|^2|2Y_{,\zz}|^2\longrightarrow
2m{\rm Re}(\rho )-e^2\rho\bar\rho$, without changing the congruence.

The charged solution was given in Newman et al.\ (1965). They claimed that
the metric can be generated by a classical charged rotating ring. As in
the previous paper Newman and Janis (1965), it was then admitted in a
footnote that the reason why this cannot be true had already been
explained to them.

\section{Appendix: Standard Notation}

Let $\{\be_a\}$ and $\{\oo^a\}$ be dual bases for tangent vectors
and linear 1-forms, respectively, i.e., $\oo^a( \be_b)=\dd^a_b$.
Also let $g_{ab}$ be the components of the metric tensor,
\beqn
ds^2 = g_{ab}\oo^a\oo^a, \quad g_{ab} = \be_a\cdot\be_b.
\eeqn
The components of the connection in this frame are the Ricci
rotation coefficients,
$$
{\CC^a}_{bc} = -
{\oo^a}_{\mu;\nu}{e_b}^\mu {e_c}^\nu, \qquad
\CC_{abc} = g_{as}{\CC^s}_{bc},
$$
The commutator coefficients ${D^a}_{bc} = - {D^a}_{cb}$ are
defined by
\beqn
[\be_b, \be_c] = {D^a}_{bc}\be_a, \quad \rm{where}\quad
[\bu,\bv](f) = \bu(\bv(f)) - \bv(\bu(f)).
\eeqn
or equivalently by
\beq\label{E:doo}
d \oo^a = {D^a}_{bc}\oo^b\wedge\oo^c.
\eeq

Since the connection is symmetric, ${D^a}_{bc} = - 2{\CC^a}_{[
bc]}$, and since it is metrical
\beqn
\CC_{abc} = \half (g_{ab|c} + g_{ac|b} - g_{bc|a}
+ D_{bac} + D_{cab} - D_{abc}),
\eeqn
\beqn
\CC_{abc} = g_{am}{\CC^m}_{bc},\qquad D_{abc}= g_{am}{D^m}_{bc}.
\eeqn
If it is assumed that the $g_{ab}$ are constant, then the connection
components are determined solely by the commutator coefficients and
therefore by the exterior derivatives of the tetrad vectors,
\beqn
\CC_{abc} = \half (D_{bac} + D_{cab} - D_{abc}).
\eeqn

The components of the curvature tensor are
\beq\label{E:riem}
{\Theta^a}_{bcd} \equiv {\CC^a}_{bd|c} - {\CC^a}_{bc|d}
+ {\CC^e}_{bd}{\CC^a}_{ec} - {\CC^e}_{bc}{\CC^a}_{ed}
- {D^e}_{cd}{\CC^a}_{be}.
\eeq
We must distinguish between the expressions on the right, the
${\Theta^a}_{bcd}$, and the curvature components, ${R^a}_{bcd}$,
which the N--P formalism treat as extra variables, their ($\Psi_i$).

A crucial factor in the discovery of the spinning black hole
solutions was the use of differential forms and the Cartan
equations. The connection 1-forms ${\Gb^a}_b$ are defined as
\beqn
{\Gb^a}_b = {\CC^a}_{bc} \oo^c.
\eeqn
These are skew-symmetric when $g_{ab|c} = 0$,
\beqn\label{}
\Gb _{ba} = - \Gb_{ab}, \quad \Gb_{ab} = g_{ac}{\Gb^c}_a.
\eeqn
The first Cartan equation follows from (\ref{E:doo}),
\beq\label{E:cart:a}
d \oo^a + {\Gb^a}_b\oo^b = 0.
\eeq

The curvature 2-forms are defined from the second Cartan equations,
\beq\label{E:cart:b}
{\Tb^a}_b \equiv d{\Gb^a}_b + {\Gb^a}_c \wedge {\Gb^c}_b
= \half {R^a}_{bcd}\oo^c\oo^d.
\eeq
The exterior derivative of (\ref{E:cart:a}) gives
\beqn
{\Tb^a}_b\wedge \oo^b = 0\quad \Rightarrow \quad {\Theta^a}_{[bcd]} =
0,
\eeqn
which is just the triple identity for the Riemann tensor,
\beq\label{triple}
{R^a}_{[bcd]} = 0.
\eeq
Similarly, from the exterior derivative of \eqref{E:cart:b},
\beqn
d{\Tb^a}_b - {\Tb^a}_f \wedge {\Gb^f}_b + {\Gb^a}_f \wedge
{\Tb^f}_b = 0,
\eeqn
that is
\beqn
{\Tb^a}_{b[cd;e]} \equiv 0, \ra {R^a}_{b[cd;e]} = 0.
\eeqn
This equation says nothing about the Riemann tensor, ${R^a}_{bcd}$
directly. It says that certain combinations of the derivatives of
the expressions on the right hand side of (\eqref{E:riem}) are
linear combinations of these same expressions.
\begin{gather}\label{E:bianchi}
\Theta_{ab[cd|e]} + {D^s}_{[cd} \Theta_{e]sab}
- {\CC^s}_{a[c} \Theta_{de]sb}- {\CC^s}_{b[c} \Theta_{de]as} \equiv 0.
\end{gather}
These are the true Bianchi identities. A consequence of this is
that if the components of the Riemann tensor are thought of as
variables, along with the components of the metric and the base
forms, then these variables have to satisfy
\beq\label{E:bianchiR}
R_{ab[cd|e]} = -2R^a_{be[c}\CC^e_{df]}.
\eeq

\end{document}